\documentclass[10pt,a4paper,onecolumn]{article}                
\usepackage{latexsym}
\usepackage[a4paper,left=2.5cm,right=2.5cm,top=2.5cm,bottom=2.5cm]{geometry}
\usepackage{amsmath}
\usepackage{amsfonts}
\usepackage[pagebackref]{hyperref}
\begin{document}

\title{What is the limit $\hbar \to 0$ of quantum theory?}

\author{U. Klein\thanks{ulf.klein@jku.at}\\University of Linz\\Institute for Theoretical Physics\\ A-4040 Linz, Austria\\}

\date{\today}

\maketitle
\begin{center}
American  Journal of Physics, in press  
\end{center}

\begin{abstract}
An analysis is made of the relation between quantum theory and classical mechanics, in the 
context of the limit  $\hbar \to 0$. Several ways in which this limit may be performed are
considered. It is shown that Schr\"odinger's equation for a single particle moving in an 
external potential $V$  does not, except in special cases, lead, in this limit, to Newton's 
equation of motion for the particle. 
This shows that classical mechanics cannot be regarded 
as emerging from quantum mechanics---at least in this
sense---upon straightforward application of the limit $\hbar \to 0$.
\end{abstract}

\maketitle

\section{Introduction}
\label{sec:Introduction}

Dirac's famous book~\cite{dirac:principles_p88} on quantum theory 
states that  ``...classical mechanics may be regarded as the limiting 
case of quantum mechanics when  $\hbar$ tends to zero.'' In quantum 
mechanics a single particle in an external potential is described by 
Schr\"odinger's equation,    
\begin{equation}
  \label{eq:SPSCHROE}
\bigg[ \frac{\hbar}{i} \frac{\partial}{\partial t}
- \frac{\hbar^{2}}{2 m }
\sum_{k=1}^{3} \left( \frac{\partial}{\partial x_k} \right)^{2}
+V(x,t) 
\bigg]\psi(x,t)=0 
\mbox{.}
\end{equation} 
Thus, Dirac's statement would imply that Newton's second law, 
\begin{equation}
\label{eq:NSEDSCHL}
\frac{d}{dt} m r_k(t)  =  p_k(t),
\hspace{0.5cm}
\frac{d}{dt} p_k(t)  = -\frac{\partial V(x,t)}{\partial x_k}\Big|_{x=r(t)},
\end{equation}
should follow from Schr\"odinger's equation in the limit $\hbar \to 0$. 

Nobody has ever 
performed a general exact calculation showing that Eq.~\eqref{eq:SPSCHROE} 
implies Eq.~\eqref{eq:NSEDSCHL} in the limit $\hbar \to 0$. This 
conflict leads frequently to the statement that this limit cannot be understood in such 
a simple way. But this answer is not really satisfying. The attitude of the scientific 
community with regard to this point---which is extremely important 
for the interpretation of quantum theory (QT) as well as for more general questions such as 
the problem of reductionism---is somewhat inconsistent. On the one hand, Dirac's dictum---which 
has been approved by other  great physicists---is considered to be true. On the other hand it cannot 
be verified. Since the beginnings of QT, a never-ending series of works deal with this question, but 
the deterministic limit of QT, in the sense of the above general statement, has 
never been attained. Frequently, isolated ``classical properties'' which indicate 
asymptotic ``nearness'' of deterministic physics, or structural similarities (such as those between 
Poisson brackets and commutators) are considered as  a 
substitute for the limit. The point is that in most of these papers 
(see e.g.~\cite{rowe:classical,werner:classical,allori_zanghi:classical} 
to mention only a few) 
the question ``what is the limit $\hbar \to 0$ of quantum theory?'' is 
\emph{not} studied. It is taken for granted that the final answer to this question 
has already been given (by Dirac and others) and the remaining problem is just 
how to confirm or illustrate it by concrete calculations. But none of these attempts 
is satisfying. 

In the present paper the question formulated in the title will be studied without knowing 
the answer. A detailed step-by-step style of exposition has been chosen in order to 
understand this singular limit. In 
fact, the paper has been written with the idea in mind to provide an in-depth answer to 
a student's  question concerning 
the mathematical details of the transition from Eq.~(\ref{eq:SPSCHROE}) 
to Eq.~(\ref{eq:NSEDSCHL}). 
The questions how to perform a limit and how to characterize the relations 
between different (related) physical theories are closely connected to the basic question of 
how to characterize physical theories themselves. We take a pragmatic position with respect to this 
question and characterize a physical theory simply by the set of its 
predictions. This leads automatically to a 
reasonable definition of limit relations between different physical theories. 

We start by 
discussing,  in Section~\ref{sec:two-exampl-class}, two well-understood concrete limiting 
relations between two pairs of classical physical theories. These classical 
limiting relations, referred to as \emph{standard limit} and \emph{deterministic limit},
define possible meanings of the phrase ``the limit $\hbar \to 0$'' in QT.  
The term ``deterministic limit'' is closely related to the notion of a 
``deterministic theory'' used in this paper. The latter is (in contrast 
to an ``indeterministic theory'') able to make deterministic predictions 
(with probability~1) on single events. In other words, the predictions 
of a deterministic theory can be verified in single experiments, 
in contrast to experiments on statistical ensembles, which are required to verify
predictions of an indeterministic theory. In this context the term 
probabilistic is a synonym for indeterministic. Note that an indeterministic 
theory is, as a rule, based on deterministic equations, i.e., on equations whose solutions at a time $t$ are uniquely determined by initial values at 
an earlier time $t_0$. It is the physical meaning of the variables that 
makes a theory deterministic or indeterministic.

In Section~\ref{sec:stand-limit-quant} we use the variables introduced by Madelung 
to obtain the standard limit of QT, previously found by Rosen,\cite{rosen:classical_quantum} 
Schiller,\cite{schiller:quasiclassical} and others. In Section~\ref{sec:formal_standard-limit-quant} 
we derive the deterministic limit of the standard limit of QT. We find that Ehrenfest's relations, 
which have not been taken into account in previous 
treatments,\cite{rosen:classical_quantum,cohn:quantum,kobe:comments,nikolic:classical,gondran:discerned} provide the missing link between the standard limit (field) 
theory and the trajectory equations of Newtonian mechanics (NM). In 
Section~\ref{sec:formal-limit-quant} we investigate the deterministic limit of QT and 
conclude that this limit does not exist. In Section~\ref{sec:combi-limit-quant} we try to 
reconstruct the states of NM from QT by combining the deterministic limit and the 
standard limit. In this way we are indeed able to identify a class of (three) potentials and 
states that allow for a transition from QT to NM in the limit $\hbar \to 0$.  These include 
among others the coherent states of the harmonic oscillator.\cite{schroedinger:continuous} 
In Section~\ref{sec:are-all-potentials} we try to extend this process to arbitrary potentials. 
The obtained results are discussed and interpreted in Section~\ref{sec:discussion} and the 
final conclusion is drawn in Section~\ref{sec:conclusion}.

\section[Two examples of classical limit relations]{Two  examples of limit 
relations in classical physics}
\label{sec:two-exampl-class}

As our first example we consider the relation between relativistic mechanics 
and NM. As is well known, in relativistic mechanics a new fundamental constant, the speed of light 
$c$, appears, which is absent (infinitely large) in NM. Otherwise, the mathematical 
structures of both theories are similar. The basic equations of relativistic mechanics differ 
from Eqs.~(\ref{eq:NSEDSCHL}) only by factors of $\gamma$, which depend on $v/c$ and disappear 
(reduce to~1) if $c$ becomes large:
\begin{equation}
  \label{eq:THLOFSG}
\gamma=\sqrt{1-\frac{v^{2}}{c^{2}}},\hspace{1cm} 
\lim_{c \rightarrow \infty} \gamma= 1
\mbox{.}
\end{equation}
The relation between relativistic mechanics and NM may be summarized as follows: 
\begin{itemize}
\item Both theories have the same mathematical structure: ordinary 
differential equations for trajectories. A new fundamental constant $c$  
appears in relativistic mechanics.
\item The limit $1/c \rightarrow 0$ transforms the basic equations of relativistic 
mechanics into the basic equations of NM; the same is true for the solutions of 
these equations.
\end{itemize}
We see that relativistic mechanics and NM provide a perfect realization of a limit relation 
(NM is the limit theory of relativistic mechanics) or a covering relation (relativistic 
mechanics is the covering theory of NM). The significant feature is the 
appearance of a new fundamental constant which allows for a transition between two 
different theories of the same mathematical type. We will refer to the type of limit 
relation encountered in this first example as \emph{standard} limit relation.

Our second example concerns the relation between NM and a probabilistic 
version of NM, which can be constructed according to the following well-known 
recipe. We consider a phase space probability density $\rho(x,p,t)$ and assume 
that the total differential of $\rho$ vanishes:
\begin{equation}
  \label{eq:WEAT23TTD}
d\rho(x,p,t)=
\frac{\partial \rho}{\partial x_k} dx_k+
\frac{\partial \rho}{\partial p_k} dp_k+
\frac{\partial \rho}{\partial t} dt =0
\mbox{.}
\end{equation}
This means that $\rho$ is assumed to be constant along arbitrary infinitesimal 
changes of $x_k,\,p_k,\,t$. Next we postulate that the movement in phase space 
follows classical mechanics, i.e., we set $dx_k=(p_k/m)dt$ and 
$dp_k=-(\partial V/\partial x_k)dt$. This leads to the 
partial differential equation (Liouville equation)  
\begin{equation}
  \label{eq:AI3SP7DFNM}
\frac{\partial \rho}{\partial t}+
\frac{p_k}{m} \frac{\partial \rho}{\partial x_k}
-\frac{\partial V}{\partial x_k}\frac{\partial \rho}{\partial p_k} =0
\mbox{,}
\end{equation}
which has to be solved by choosing initial values $\rho(x,p,0)$ for 
the new dynamical variable $\rho(x,p,t)$.  The relation between the probabilistic 
version (of NM) and NM may be summarized as follows: 
\begin{itemize}
\item The probabilistic version and NM have different mathematical structures; the 
probabilistic version is ruled by  a partial differential equation, NM by an ordinary 
differential equation. No new constant appears in the probabilistic version.
\item The probabilistic version and NM belong to fundamentally different epistemological categories. 
NM is a deterministic theory. The probabilistic version is a probabilistic (indeterministic) 
theory; predictions about individual events cannot be made because the initial values 
for individual particles are unknown. 
\end{itemize}
The absence of a new fundamental constant prevents a simple transition between 
the two theories as found in our first example. Nevertheless, a kind of 
limit relation can be established by means of appropriate (singular) 
\emph{initial values}. A probability density that is sharply peaked 
at $t=0$ retains its shape at later times. Inserting the Ansatz
\begin{equation}
  \label{eq:IM9ESA2HON}
\rho(x,p,t) = \delta^{(3)}(x-r(t))\,\delta^{(3)}(p-p(t))
\end{equation}
into Eq.~(\ref{eq:AI3SP7DFNM}), it is easily shown
that admissible particle trajectories $r_k(t)$, $p_k(t)$  are just 
given by the solutions of Newton's equations~(\ref{eq:NSEDSCHL}). 
Thus, NM can be considered as a limit theory of the probabilistic version in the 
sense that the manifold of solutions of a properly (with 
regard to singular initial values) generalized version of the probabilistic version 
leads to NM. This limit relation is weaker than the one 
encountered in our first example, because there is no mapping of 
individual solutions. By allowing for singular solutions we have essentially 
constructed the \emph{union} of the deterministic theory NM 
and the original probabilistic version of NM; it is then no surprise 
that the generalized probabilistic version theory contains NM as a special case. 
Considered from a formal point of view, however, the (generalized) probabilistic 
version is a perfect covering theory since its manifold of solutions is larger than 
that of NM. We shall refer to the kind of limit relation found in this second 
example as a \emph{deterministic} limit relation.

\section[The standard limit of quantum theory]{The standard limit of quantum 
theory}
\label{sec:stand-limit-quant}

Let us now compare QT and NM [Eqs.~\eqref{eq:SPSCHROE} and~\eqref{eq:NSEDSCHL}] in the 
light of the above examples. The two theories obviously differ in their mathematical 
structures; this indicates the possibility to obtain NM from QT by means of a deterministic limiting 
process. However, in addition, a new fundamental constant ($\hbar$) appears in 
QT; this indicates the possibility to obtain NM from QT by means of a standard limiting process. 
Thus, the limiting process that leads from QT to NM is either nonexistent or more 
complex than either of the above examples.

Let us try, as a first step, to perform the standard limit of QT, as defined by the first example 
in Section~\ref{sec:two-exampl-class}. Taking the limit $\hbar \to 0$  in 
Eq.~\eqref{eq:SPSCHROE} produces a nonsensical result. This indicates that the 
real and imaginary parts of $\psi$ are not appropriate variables with regard to this 
limiting process, probably because they become singular in the 
limit $\hbar \to 0$. Thus, a different set of dynamical variables, with regular behavior in the limit, should be 
chosen. There is convincing evidence, 
from various physical contexts, that appropriate 
variables, denoted by $\rho$ and $S$, are defined by the transformation
\begin{equation}
  \label{eq:HDUI9J2UI}
\psi = \sqrt{\rho}\,\mathrm{e}^{i S/\hbar}
\mbox{.}
\end{equation}
This transformation has been introduced by Madelung.~\cite{madelung:quantentheorie}
Note that using these variables in a meaningful limiting process requires that the modulus of 
$\psi$ remains regular while its phase diverges like $\hbar^{-1}$ for small $\hbar$. This 
singular behavior, which was noted very early,~\cite{vanvleck:correspondence} is the behavior 
of the majority of ``well-behaved'' quantum states. Other, more singular, states may, however, behave 
in a different manner and will then require a different factorization in terms of $\hbar$. This 
may also lead to different equations in the limit $\hbar \to 0$; an example will be given in 
Section~\ref{sec:combi-limit-quant}.

In terms of the new variables, Schr\"odinger's equation takes the form of 
two coupled nonlinear differential equations. The first is a continuity equation 
which does not contain  $\hbar$,
\begin{equation}
 \label{eq:CA2IH3TMF}
\frac{\partial \rho}{\partial t}+\frac{\partial}{\partial x_{k}}
\frac{\rho}{m}  \frac{\partial S}{\partial x_{k}}=0
\mbox{.}   
\end{equation} 
The second equation contains $\hbar$ as a proportionality 
factor in front of a single term:
\begin{equation}
\label{eq:QHKUZ24MF}
\frac{\partial S}{\partial t}+ \frac{1}{2m}
\sum_{k}
\left( \frac{\partial S}{\partial x_{k}} \right)^{2} +V  -
\frac{\hbar^2}{2m}\frac{\triangle\sqrt{\rho}}{\sqrt{\rho}}
=0
\mbox{.} 
\end{equation}
Equation~\eqref{eq:QHKUZ24MF} is referred to as quantum Hamilton-Jacobi 
equation (QHJ). The $\hbar$-dependent ``quantum term'' in Eq.~\eqref{eq:QHKUZ24MF} 
describes the influence of $\rho$ on  $S$. (It is frequently denoted as 
``quantum potential,'' which is an extremely misleading nomenclature 
because a potential is, as a rule, an externally controlled quantity.) Its 
physical meaning, as interpreted by the present author, has been  
discussed in more detail elsewhere.~\cite{klein:schroedingers}
  
In the limit $\hbar \to 0$ the quantum term disappears. Thus the 
standard limit of QT is given by two partial differential equations: the continuity 
equation~\eqref{eq:CA2IH3TMF}, which depends on $\rho$ and $S$, and the 
Hamilton-Jacobi (HJ) equation,
\begin{equation}
\label{eq:QHCL14MF}
\frac{\partial S}{\partial t}+ \frac{1}{2m}
\sum_{k}
\left( \frac{\partial S}{\partial x_{k}} \right)^{2} +V  
=0
\mbox{,} 
\end{equation} 
which depends only on $S$. The two equations~\eqref{eq:CA2IH3TMF} 
and~\eqref{eq:QHCL14MF}, which will be referred to as probabilistic 
Hamilton-Jacobi theory (PHJ), constitute the classical limit of Schr\"odinger's 
equation or single-particle QT. Clearly, this limit does not 
agree with the trajectory equations~\eqref{eq:NSEDSCHL} of NM.

Much confusion has been created by the fact that the Hamilton-Jacobi 
formulation of classical mechanics allows the determination of particle 
trajectories with the help of the HJ equation. From the fact that this 
equation can be obtained from QT in the limit $\hbar \to 0$ it is often 
concluded, neglecting the continuity equation, that classical mechanics is 
the $\hbar \to 0$ limit of QT. However, the limit $\hbar \to 0$ of QT does 
not provide us with the theory of canonical transformations, which is 
required to actually calculate particle trajectories. Note also that for exactly 
those quantum-mechanical states that are most similar to classical states 
(e.g., coherent states, see Section~\ref{sec:combi-limit-quant}), the 
classical limit of QHJ \emph{differs} from HJ. There \emph{is} in fact a 
connection between the PHJ and NM, but this requires a second limiting process, 
as will be explained in Section~\ref{sec:formal_standard-limit-quant}.

Both the PHJ and its (standard) covering theory QT are probabilistic 
theories, which provide statistical predictions (probabilities and 
expectation values) if initial values for $S$ and $\rho$ are specified.
Although we now have partial differential equations, the relation 
between QT and PHJ resembles in essential aspects the relation 
between relativistic mechanics and NM.

\section[The deterministic limit of the standard limit of quantum theory]{The 
deterministic  limit of the standard limit of quantum theory}
\label{sec:formal_standard-limit-quant}

The deterministic limit of the classical limit PHJ of QT is of considerable interest for the 
present problem, despite the fact that the PHJ no longer contains $\hbar$. Existence 
of a deterministic limit implies that $\rho(x,t)$ takes the form of a delta function peaked 
at trajectory coordinates $r_k(t)$ [which, hopefully, should then be solutions of the classical 
Eqs.~(\ref{eq:NSEDSCHL})]. Thus, adopting a standard formula, we may write  $\rho(x,t)$ 
as an analytic function   
\begin{equation}
  \label{eq:AFT18FO6TD}
\rho_{\epsilon}(x,t)= \left(\frac{1}{\pi \epsilon} \right)^{3/2}
\exp \left\{-\frac{1}{\epsilon}\sum_{k=1}^{3}\left[x_k-r_k(t) \right]^{2} \right\},
\end{equation}
which represents $\delta^{(3)}(x-r(t))$ under the integral sign in the 
limit $\epsilon \to 0$:
\begin{equation}
  \label{eq:D7HLMM9F}
\lim_{\epsilon \rightarrow 0}\rho_{\epsilon}(x,t)=\delta^{(3)}(x-r(t))
\mbox{.}
\end{equation}
In order to check whether or not this deterministic representation of 
$\rho(x,t)$ is compatible with the basic equations of PHJ, we 
insert the Ansatz~(\ref{eq:AFT18FO6TD}) into the continuity 
equation~(\ref{eq:CA2IH3TMF}) and calculate the derivatives. After some
rearrangement, Eq.~(\ref{eq:CA2IH3TMF}) takes the form~\cite{nikolic:classical}   
\begin{equation}
  \label{eq:EZ4SRW8MGB}
\rho_{\epsilon}(x,t)
\left\{
\left[x_k-r_k(t)\right]
\left(p_k(t)-\frac{\partial S(x,t)}{\partial x_k}\right) 
+ \frac{\epsilon}{2} 
\frac{\partial}{\partial x_k} \frac{\partial}{\partial x_k} S 
\right\}=0
\mbox{,}
\end{equation}
where $p_k(t)=m\dot{v}_k(t)$. At this point we recall that in the PHJ a 
momentum field $p_k(x,t)$, defined by  
\begin{equation}
  \label{eq:JU25ZT7SD}
p_k(x,t)=\frac{\partial S(x,t)}{\partial x_k}
\end{equation}
exists. The trajectory momentum $p_k(t)$ should be clearly distinguished from 
this field momentum $p_k(x,t)$. 

In the limit $\epsilon \to 0$, $\rho_{\epsilon}$ becomes a distribution 
and both sides of Eq.~(\ref{eq:EZ4SRW8MGB}) have to be integrated 
over three-dimensional space in order to obtain a mathematically well-defined 
expression. The first term in the bracket vanishes as a consequence of the term 
$x_k-r_k(t)$ (at this point we start to disagree with Ref.~\cite{nikolic:classical}). 
The second term vanishes too for $\epsilon \to 0$ provided the second derivative 
of $S$ is regular at $\epsilon=0$. But this can safely be assumed since the 
equation for $S$ does not contain $\rho$. We conclude that the singular (deterministic) 
Ansatz~(\ref{eq:AFT18FO6TD}) is a valid solution of PHJ for arbitrary~$S$. 

The present theory is incomplete since differential equations 
for the particle trajectories $r_k(t)$ still have to be found. 
Generally, two conditions must be fulfilled in order to define particle 
coordinates in a probabilistic theory, namely (i) the limit of sharp 
(deterministic) probability distributions must be a valid solution, and 
(ii) an evolution law for the time-dependent expectation values must exist.  
We have just shown that the first (more critical) condition is fulfilled; Ehrenfest-like 
relations corresponding to the second condition exist in almost all statistical 
theories. For the PHJ these take exactly the same form as in QT, 
namely~\cite{kocis:ehrenfest,klein:statistical}
\begin{eqnarray}
\frac{d}{dt} \overline{x_k} & = & \frac{\overline{p_k}}{m}
\mbox{,}
\label{eq:FIRUZT2TSCHL}\\
\frac{d}{dt} \overline{p_k} & = &
\overline{-\frac{\partial V(x,t)}{\partial x_k}}
\label{eq:SEUZT3TSCHL}
\mbox{,}
\end{eqnarray}
where average values such as $\overline{x_k}$ are defined according to the standard expression
\begin{equation}
  \label{eq:DE1HFDZ8EV}
\overline{x_k}(t) =\int \mathrm{d^{3}} x\, \rho(x,t)\, x_k
\mbox{.}
\end{equation}
From Eq.~\eqref{eq:FIRUZT2TSCHL} and the continuity equation~\eqref{eq:CA2IH3TMF} 
we obtain the useful relation
\begin{equation}
  \label{eq:AZGR345FFDE}
\overline{p_k}(t)  = \int \mathrm{d^{3}} x\, \rho(x,t)\, 
\frac{\partial S(x,t)}{\partial x_k}
\mbox{.}
\end{equation}
Since we have shown that the deterministic limit for $\rho$ is a valid solution of PHJ, we may 
now use Eq.~\eqref{eq:D7HLMM9F} and obtain in the limit $\epsilon \to 0$ the 
identification of trajectory quantities, 
\begin{equation}
  \label{eq:FI21OTR89QU}
\overline{x_k}(t) = r_k(t), \qquad \overline{p_k}(t) = m\dot{r}_k(t) =p_k(t) 
\mbox{,}
\end{equation}
from the definitions of the expectation values. The differential relations connecting these 
quantities follow from Ehrenfest's theorem and agree with the basic 
Eqs.~\eqref{eq:NSEDSCHL} of NM.  A completely different type of physical law 
has emerged from the field theoretic relations of the PHJ theory. Thus, classical 
mechanics is, indeed, contained in PHJ as a deterministic limit, in analogy to the 
second example of Section~\ref{sec:two-exampl-class}.  

Equation~\eqref{eq:AZGR345FFDE} takes in this limit the form
\begin{equation}
  \label{eq:TRA12TRAQU}
p_k(t)=\frac{\partial S(x,t)}{\partial x_k}\Big|_{x=r(t)}=\frac{\partial S(r(t),t)}{\partial r_k(t)}
\mbox{,}
\end{equation}
which provides an interesting link between a particle variable and a field variable. 
We expect for consistency that this link admits a derivation of the equation for $\dot{p}$ 
[see Eqs.~\eqref{eq:NSEDSCHL}] from the (field-theoretic) HJ~equation.  This is indeed the case. 
We calculate the derivative of the HJ equation~\eqref{eq:QHCL14MF} with respect to $x_i$, 
change the order of derivatives with respect to $x_i$ and $t$, and project the resulting
relation on the trajectory points $x_k=r_k(t)$. This leads to the equation
\begin{equation}
  \label{eq:NEI8LTRS1F}
\frac{\partial}{\partial t}\Big|_{2.}
\frac{\partial S(r(t),t)}{\partial r_i(t)} +
\frac{1}{m} \frac{\partial S(r(t),t)}{\partial r_k(t)}
\frac{\partial^{2} S(r(t),t)}{\partial r_i(t) \partial r_k(t)}+
\frac{\partial V(r(t),t)}{\partial r_i(t)} =0
\mbox{, }
\end{equation}
where the notation indicates that the time derivative operates on the second argument 
of $S$ only. Using now Eq.~\eqref{eq:TRA12TRAQU} and the definition of particle momentum, 
we see that the first two terms of Eq.~\eqref{eq:NEI8LTRS1F}  agree exactly with the (total) 
time derivative of $p_i(t)$ and Eq.~\eqref{eq:NEI8LTRS1F} becomes the second Newton equation. 
This establishes the connection between the PHJ equations and trajectory differential equations 
mentioned in Section~\ref{sec:stand-limit-quant}, and completes our treatment of the 
deterministic limit of the PHJ theory. This derivation of NM seems to be new; it is based 
on several interesting 
papers~\cite{rosen:classical_quantum,cohn:quantum,kobe:comments,nikolic:classical,
gondran:discerned} reporting important steps in the right direction.

\section[The deterministic limit of quantum theory]{The deterministic  limit of quantum theory}
\label{sec:formal-limit-quant}

In QT, the coupling term  in QHJ [see Eq.~\eqref{eq:QHKUZ24MF}] prevents a deterministic limit
of the kind found for the PHJ. To see this, we start from the assumption that a quantum 
mechanical system exists that admits a solution of the 
form of Eq.~\eqref{eq:AFT18FO6TD} for arbitrary $t$. Inserting Eq.~\eqref{eq:AFT18FO6TD} 
into Eqs.~\eqref{eq:CA2IH3TMF} and \eqref{eq:QHKUZ24MF} leads to two equations. The first 
is the continuity equation, which takes the same form~\eqref{eq:EZ4SRW8MGB} as 
before. The second is the QHJ, which takes the form
\begin{equation}
\label{eq:QHIMIR23MF}
\frac{\partial S}{\partial t}+ \frac{1}{2m}
\sum_{k}
\left( \frac{\partial S}{\partial x_{k}} \right)^{2} +V  +
\frac{1}{2m}
\left(  \frac{\hbar}{\epsilon} \right)^{2} 
\left\{ 3 \epsilon- \sum_{k=1}^{3} \left[x_k-r_k(t) \right]^{2}\right\}
=0
\mbox{.} 
\end{equation}  
Equation~\eqref{eq:QHIMIR23MF} shows that the coupling term diverges (for finite $\hbar$) in the 
limit $\epsilon \to 0$. Consequently, there is no reason to expect that the second derivative 
of $S$ with respect to $x_k$  [see Eq.~\eqref{eq:EZ4SRW8MGB}]  is regular at $\epsilon \to 0$ and 
that a delta-function-like $\rho(x,t)$, as given by  Eq.~\eqref{eq:AFT18FO6TD}, can be a solution 
of Eqs.~\eqref{eq:CA2IH3TMF} and~\eqref{eq:QHKUZ24MF}. Thus, the deterministic limit of QT (if it exists) 
cannot be obtained in the same way as in the PHJ. 

We next consider several concrete solutions of QT that lead to probability densities 
similar to Eq.~\eqref{eq:AFT18FO6TD}. As a first example we consider an ensemble 
of free particles that are distributed at $t=0$ according to a probability 
density~\eqref{eq:AFT18FO6TD} centered at $r_k(0)=0$ [set $r_k(t)=0$ 
in Eq.~\eqref{eq:AFT18FO6TD}]. The initial value for $S(x,t)$ is given by $S(x,0)=p_{0,k}x_k$, 
i.e.,  $S(x,t)$ fulfills at $t=0$ the deterministic relation~\eqref{eq:TRA12TRAQU}. These 
initial values describe for small $\epsilon$ a localized, classical particle 
in the sense that there is no uncertainty with respect to position or momentum. 
A calculation found in many textbooks leads to the following solution of 
Schr\"odinger's equation for $\rho$:
\begin{equation}
\rho(x,t)= \left(\frac{1}{\pi A(t)} \right)^{\frac{3}{2}}
\exp \left\{-\frac{1}{A(t)}\sum_{k=1}^{3}
\left[ x_k - r_k(t)  \right]^{2} \right\},
\label{eq:ALG5SC9IDON}
\end{equation}
where $mr_k(t)=p_{0,k}t$ and $A(t)=A_{f}(t)=\epsilon [1+(\hbar/\epsilon)^{2}(t/m)^{2} ]$. 
We see from Eq.~\eqref{eq:ALG5SC9IDON} that the peak of $\rho$ moves in agreement with
NM, but the width of the wave packet increases with increasing time as well as with 
\emph{decreasing} $\epsilon$.  A complete localization can be achieved only at $t=0$.
At later times the quantum uncertainty, due to the finite $\hbar$, dominates the behavior 
of the ensemble completely, despite our choice of ``deterministic'' initial conditions. 
      
As a second example, we consider an ensemble of particles moving in a harmonic oscillator
potential $V(x)=(m\omega^{2}/2)x_kx_k$, using exactly the same initial conditions as in the
above example of force-free motion. The result for $\rho$ takes the same 
form as for the force-free ensemble [see Eq.~\eqref{eq:ALG5SC9IDON}], but 
with 
$mr_k(t)=(p_{0,k}/\omega) \sin \omega t$ and 
\begin{equation}
  \label{eq:JW2SD6LH9Z}
A(t)=A_{h}(t)=\epsilon 
\left[ \cos^{2}\omega t + 
\frac{1}{m^{2}\omega^{2}} \left( \frac{\hbar}{\epsilon}\right)^{2}\sin^{2}\omega t \right].
\end{equation}
The width $A_{h}(t)$ increases again with decreasing $\epsilon$ and prevents again a deterministic 
limit. We mention, without going into details,~\cite{ter_haar:problems} that a third example showing 
the same behavior can be found, namely an ensemble of particles moving under the influence of a 
constant force. 

The three examples considered in this section correspond to three potentials proportional 
to $x_k^{n}$, where $n=0,1,2$. For these  potentials the expectation values of the 
corresponding forces fulfill the relation $\overline{F_k(x)}=F_k(\overline{x})$. Therefore, equations 
of motion for $\overline{x_k}$ and $\overline{p_k}$ exist as a consequence of Ehrenfest's theorem.
Despite these classical features, even these ``optimal'' states do not permit 
a deterministic limit of QT. We conclude, in agreement with common wisdom, 
that this limit does not exist.

\section[The combined limit of quantum theory]{The combined limit of quantum theory}
\label{sec:combi-limit-quant}

Let us summarize what has been achieved so far. In Section~\ref{sec:stand-limit-quant} 
the limit $\hbar \to 0$ has been taken for arbitrary quantum states (including wave 
packets with fixed width $\epsilon$). The result of this first ``standard limiting process'' was a 
classical statistical theory referred to as probabilistic Hamilton-Jacobi 
theory (PHJ). In Section~\ref{sec:formal_standard-limit-quant} 
the limit $\epsilon \to 0$ of PHJ has been taken. The result of this 
second ``deterministic limit'' was Newtonian mechanics (NM). Therefore NM is a subset of the classical limit PHJ of quantum theory (QT),
but NM is not \emph{the} classical limit of QT, since we cannot neglect almost all of the (statistical) 
states of PHJ.  Thus, the two limiting processes performed in this order have not led us from
QT to NM in the sense that NM can be said to be the classical limit $\hbar \to 0$ of QT. 
In Section~\ref{sec:formal-limit-quant} it has been shown that inverting the order of the two 
limiting processes (first $\epsilon \to 0$ then $\hbar \to 0$) does not solve the problem either 
since the limit $\epsilon \to 0$  (with $\hbar$ fixed)  does not exist. The two limiting processes 
clearly do not commute. Thus, it is impossible to obtain NM as the classical limit of QT, no 
matter which order of the two (separate) limiting processes is chosen.

Fortunately, we have still the option to \emph{combine} both 
limits.  That is, we could assume that the width of the wave packets is a 
monotonic function of $\hbar$. This means that the localization of wave 
packets (the deterministic limit) and the change of the basic equations of QT  
(the standard limit) take place 
\emph{simultaneously} in the limit $\hbar \to 0$. Such states seem 
artificial from the point of view of experimental verification since  
the numerical value of  $\hbar$  is not under our control. Nevertheless, 
a construction of NM from QT along these lines would certainly provide 
a kind of justification for Dirac's claim that QT reduces to NM in the limit $\hbar \to 0$. Note 
also that the subject of our study is essentially of a formal nature. We are 
asking whether or not all predictions of NM can be obtained, by means of some 
limiting process $\hbar \to 0$, from the basic equations of QT. There are no 
in-principle constraints on how to perform this limit.

A brief look at  the above examples for $A(t)$ shows that  a \emph{linear} relation between 
$\hbar$ and $\epsilon$ seems most promising. Thus, we set
\begin{equation}
  \label{eq:DRMI45WWLG}
\epsilon=k\hbar
\mbox{,}
\end{equation}
were $k$ is an arbitrary constant. In order to use a notation similar 
to that in Section~\ref{sec:formal_standard-limit-quant} 
[see Eq.~\eqref{eq:AFT18FO6TD}], $\epsilon$ will be used instead of $\hbar$ as small parameter; it may be identified with  $\hbar$ in most of the 
following relations. Let us perform the identification~\eqref{eq:DRMI45WWLG} for the 
two examples considered in Section~\ref{sec:formal-limit-quant}, with 
potentials $V(x)=V_{f}(x)=0$ and 
$V(x)=V_{h}(x)=(m\omega^{2}/2)x_kx_k$, respectively. Using the same 
initial conditions as in Section~\ref{sec:formal-limit-quant}, the solutions 
for $\rho$ and $S$ of Eqs.~\eqref{eq:CA2IH3TMF} and~\eqref{eq:QHKUZ24MF} 
take essentially the same form in both cases, namely
\begin{eqnarray}
\rho(x,t)&=& \left(\frac{1}{\pi \epsilon(t)} \right)^{3/2}
\exp \left\{-\frac{1}{\epsilon(t)}\sum_{k=1}^{3}
\left[ x_k - r_k(t)  \right]^{2} \right\},
\label{eq:HZ46TRE9LPL}\\
S(x,t)&=& \frac{m}{4}\frac{\dot{\epsilon}(t)}{\epsilon(t)} 
\sum_{i=1}^{3}\left[x_i-r_i(t) \right]^{2}-\frac{1}{2}p_{k}(t)r_{k}(t)
+p_{k}(t)x_k - \sum_{i=1}^{3}\frac{\epsilon}{2k}\tan^{-1}
\frac{r_{i}(t)}{k p_{i}(t)}
\label{eq:SO8PER34DFL}
\mbox{.}
\end{eqnarray}
The solutions ($\rho_{f},\,S_{f}$)  and ($\rho_{h},\,S_{h}$)
for particle ensembles in force-free regions and linear-force fields 
may be obtained from 
Eqs.~\eqref{eq:HZ46TRE9LPL} and~\eqref{eq:SO8PER34DFL} by 
using  different widths $\epsilon(t)=\epsilon_{f}(t)$ and 
$\epsilon(t)=\epsilon_{h}(t)$, given by 
\begin{equation}
  \label{eq:JUZA2GR9WQ}
\epsilon_{f}(t) = \epsilon \left(1+\frac{t^{2}}{k^{2}m^{2}} \right), \qquad
\epsilon_{h}(t)=\epsilon 
\left[ \cos^{2}\omega t + 
\frac{1}{k^{2}m^{2}\omega^{2}} \sin^{2}\omega t \right],
\end{equation}
and different trajectory components $r_k(t)=r_k^{(f)}(t)$ and 
$r_k(t)=r_k^{(h)}$ (as well as momentum 
components $p_k(t)=m \dot{r}_k(t)$), given by 
\begin{equation}
  \label{eq:J19UHT45POI}
r_k^{(f)}(t)=\frac{p_{0,k}}{m}t , \qquad
r_k^{(h)}=\frac{p_{0,k}}{m \omega t}\sin \omega t.
\end{equation}
As Eq.~\eqref{eq:JUZA2GR9WQ} shows, both widths are 
time-dependent; for the free-particle ensemble the width increases 
quadratically, while for the bounded motion of the harmonic 
oscillator it varies periodically. However, both widths vanish in the limit
$\epsilon \to 0$ for arbitrary (finite) times $t$. This means that the 
deterministic probability density we were looking for is, in fact, 
created in this limit. The solutions for $S$ are well-behaved at 
$\epsilon=0$. The limiting process in the continuity 
equation~\eqref{eq:CA2IH3TMF} can be performed  in a similar way as 
in Section~\ref{sec:formal_standard-limit-quant} (an additional term 
due to the time dependence of $\epsilon(t)$ is regular at 
$\epsilon=0$). The remaining steps---the derivation of Newton's equations 
and their field-theoretic derivation from the QHJ---can be performed in the 
same way as in Section~\ref{sec:formal_standard-limit-quant}. Note also 
that the QHJ is regular at  $\epsilon=0$ and \emph{differs} in this 
limit from the HJ equation [cf.\ the discussion following 
Eq.~\eqref{eq:HDUI9J2UI}]. In view of a recent 
discussion~\cite{ling:limit,kazandjian:limit} it should be 
noted that this field-theoretic limit is not equivalent to its 
projection on the trajectory. 

The above solutions, with $\epsilon$ and $\hbar$ considered as 
\emph{independent} parameters (as in Section~\ref{sec:formal-limit-quant}), 
have been reported many times in the literature. It has 
also been pointed out that the special value $\epsilon=\hbar/m\omega$, 
in the harmonic oscillator example, produces the coherent states 
found by Schr\"odinger.\cite{schroedinger:continuous} On the other 
hand, the relevance of the weaker statement $\epsilon=k\hbar$ for 
the classical limit problem has apparently not been recognized. It is 
not necessary to restrict oneself to the coherent states of the harmonic 
oscillator (the special case $k=1/m\omega$) in order to obtain  
deterministic motion;  the latter may be obtained for a much larger 
class of  force-free states, harmonic oscillator states, and constant-force 
states (this last example has not been  discussed explicitly) as shown 
above. Summarizing this section, we found three potentials 
$V(x) \sim x^{n},\,n=0,1,2$, which allow for a derivation of NM from 
QT in the limit $\hbar \to 0$.  For these potentials, equations of motion for 
$\bar{x}_k,\ \bar{p}_k$ exist, as mentioned already.
Home and Sengupta~\cite{home_sengupta:} have shown that for 
these potentials the form of the quantum-mechanical solution may 
be obtained with the help of the classical Liouville theorem.

\section[Can the combined limit be performed for all potentials?]{Can the combined limit be
performed for all potentials?}
\label{sec:are-all-potentials}

We know now that three potentials exist which, for properly chosen initial 
wave packets, lead to deterministic equations of motion in the limit
$\hbar \to 0$. We shall refer to such potentials for brevity as \emph{deterministic 
potentials}. A (complete) reconstruction of NM from QT requires  that \emph{all} 
(or almost all) potentials are deterministic. In this section we ask if this can be true.

As a first point we note that the probability density $\rho(x,t)$ of all deterministic 
wave packets takes, by definition, a very specific functional form, namely one 
that reduces, like Eq.~\eqref{eq:HZ46TRE9LPL},  in the limit $\hbar \to 0$ to a delta 
function. This fixes essentially one of our two dynamic variables; we have two 
differential equations for a single unknown variable $S(x,t)$. It seems 
unlikely that this overdetermined system of equations admits solutions 
for $S(x,t)$ for \emph{arbitrary} potentials $V$.  

The second point to note is that the existence of a deterministic limit fixes 
not only the functional form of $\rho$ but also its argument. Let us assume 
that a deterministic solution for $\rho$ and $S$, with $\rho(x,t)$ taking the 
form of Eq.~\eqref{eq:HZ46TRE9LPL} with unspecified $\epsilon(t)$, exists.  The 
probability density $\rho(x,t)$ depends necessarily on $\vec{x}-\vec{r}(t)$, 
where the position vector $\vec{r}(t)$ is a solution of Newton's equation for 
the same potential $V(x)$ that occurs in the Schr\"odinger equation.  The crucial 
point is that this dependence is not created by the limiting process  $\hbar \to 0$ but 
is already present for finite $\hbar$, in the exact quantum-mechanical solution. 
For given initial conditions it has been created, so to say, by the quantum-theoretical 
formalism. This implies that  $\vec{r}(t)$ describes for finite $\hbar$ not the 
time dependence of a particle trajectory but of a position expectation value. 
Since  $\vec{r}(t)$ is (again for finite $\hbar$) the solution of Newton's equations, 
such equations for expectation values must already be present in the quantum-theoretical 
formalism. Of course, in the deterministic limit $\hbar \to 0$ the difference between particle 
trajectories and expectation values vanishes, but the important point 
is that Newton's equations must hold already for finite $\hbar$. We conclude 
that the existence of the equations of motion of NM for position expectation values 
is a necessary condition for the existence of deterministic potentials. 

This line of reasoning leads to a mathematical condition for deterministic potentials. 
Let us assume that we have a deterministic potential $V^\textrm{(det)}(x)$ in our quantum-theoretical 
($\hbar$ finite) problem. We calculate  the expectation value $\overline{x_k}(t)$ as 
defined by Eq.~\eqref{eq:DE1HFDZ8EV} using the deterministic probability 
density Eq.~\eqref{eq:HZ46TRE9LPL}. We obtain 
$\overline{x_k}(t)=r_{k}(t)$, i.e., the expectation value follows the time dependence of 
the trajectory [the peak of $\rho (x,t)$]. The latter must fulfill Newton's equation with the 
force derived from $V^\textrm{(det)}(x)$; otherwise the deterministic limit could not exist.  
Using these facts in Ehrenfest's equations, Eqs.~\eqref{eq:FIRUZT2TSCHL} and~\eqref{eq:SEUZT3TSCHL}, 
we obtain immediately the integral equation for deterministic potentials  
\begin{equation}
  \label{eq:EFU29EDIREL}
F^\textrm{(det)}_{k}(r)=\int \mathrm{d^{3}} x\,   
\delta_{\epsilon}^{(3)}(x-r)
 F^\textrm{(det)}_{k}(x)
\mbox{,}
\end{equation}
where $F^\textrm{(det)}_{k}(x)=-\partial V^\textrm{(det)}(x) / \partial x_{k}$. The quantity  $\rho(x,t)$  has been 
renamed $\delta_{\epsilon}^{(3)}(x-r(t))$ in order to show the convolution-type structure of the 
equation. Note that~Eq.~\eqref{eq:EFU29EDIREL} is only for $\epsilon > 0$ a constraint for 
$V^\textrm{(det)}(x)$. It is easy to see that a particular solution is given by $V^\textrm{(det)}(x)=a+b_kx_k+
c_{i,k}x_ix_k$, where the coefficients may depend on time. This is essentially a linear combination 
of the three deterministic potentials $x^{n},\,n=0,1,2$, found already in the last section. According 
to a theorem by Titchmarsh~\cite{titchmarsh:fourier_integrals} (a simple proof may be obtained 
with the help of the theory of generalized functions;\cite{lighthill:generalized}
see the Appendix), other solutions of Eq.~\eqref{eq:EFU29EDIREL} do not exist.  
This theorem shows that the ``combined limit'' cannot be performed for all potentials. Although 
the present treatment does not cover all conceivable physical situations, the results obtained so 
far imply already definitively that the limit $\hbar \to 0$ of QT does not generally agree with NM.

\section[Discussion]{Discussion}
\label{sec:discussion}

In our first limiting procedure, which is appropriate for ``well-behaved'' 
quantum states, we found that QT agrees in the limit $\hbar \to 0$ with a 
classical statistical theory referred to as PHJ. The latter  contains as a limiting case 
the deterministic states ruled by NM. Let us stress once again that the fact that these 
deterministic states are contained in QT does not mean that NM is the limit of QT. 
This limit is PHJ, which contains a much larger number of (probabilistic) states not 
belonging to NM. In a second attempt, the limit $\hbar \to 0$ was simultaneously 
applied to the wave-packet width and in the basic equations. We found that 
almost all states do \emph{not} admit a transition from QT to NM in the limit 
$\hbar \to 0$. 

Our final result is then that NM cannot be obtained from QT, at least by means of a mathematical limiting process  $\hbar \to 0$. This result has been 
obtained in the framework of standard QT using the Schr\"odinger picture to describe 
the quantum dynamics. One may ask whether this conclusion is specific for these choices, 
or remains true for other dynamical pictures, such as the Heisenberg picture, 
and for other formulations of QT, such as Feynman's path-integral formulation. 
Recall that the present approach is (as discussed in Section~\ref{sec:Introduction}) solely 
based on predictions, i.e., the \emph{numerical output} of the quantum theoretical 
formalism. These numbers do not depend on a particular picture of quantum dynamics.
They are also independent from the choice of a particular formulation of QT since  
all formulations of QT must lead to the same predictions. 

Let us illustrate the last point with a discussion of the path-integral formulation of 
QT.\cite{feynman.hibbs:quantum} The central quantity of this approach, the 
propagator, is an infinite sum of terms of the form $\sim\exp iS/\hbar$, 
where $S$ is the classical action and each term in the sum is to be evaluated along a 
different path between the initial and final space-time points $x_0$ and $x_1$. 
In the classical limit $\hbar \to 0$ the dominating contribution to the sum comes 
from the classical path between  $x_0$ and $x_1$, which extremizes $S$. The fact
that this path obeys the differential equations~\eqref{eq:NSEDSCHL} of NM is sometimes
interpreted in the sense of a  transition from QT to NM, which the path-integral formulation 
reveals in a particular rigourous and straighforward way. Such an interpretation is not justified. 
The form of the propagator says nothing about whether or not a particle is really 
present at the initial space-time point $x_0$; it tells us just what  will happen \emph{given that} 
a particle occupies the point $x_0$ with certainty. The second variable of QT, the probability 
density $\rho$, must also be taken into account; it enters the initial state and makes the final state 
uncertain despite the deterministic form of the propagator. A general  classification scheme 
for probabilistic theories, taking the different roles of initial values and evolution equations  
into account, has been reported recently.~\cite{klein:statistical} It is a general feature of 
classical statistical theories that the time evolution in the event space (configuration or phase 
space) is deterministic and the impossibility of making deterministic predictions (on single events) 
is solely due to uncertainty in the initial values. This classical feature is also visible in the 
phase-space theory defined by Eq.~\eqref{eq:AI3SP7DFNM} and (though in a less explicit way) in the 
configuration-space theory PHJ [see Eqs.~\eqref{eq:CA2IH3TMF} and~\eqref{eq:QHCL14MF}]. 
It is this feature, and not the transition from QT to NM, that is most explicit in the path-integral 
formalism. The classical limit of Feynman's version of QT is equivalent to the classical limit of 
Schr\"odinger's  version of QT (the PHJ), since both versions are equivalent.

Our final result is in disagreement with 
Dirac's statement quoted at the beginning of our study. Dirac discusses the problem of the classical limit 
in Section 31 of his book. He formulates the following general principle:
\begin{quote}
For any dynamical system with a classical analogue, a state for which the classical description is 
valid as an approximation is presented in quantum mechanics as a wave packet,\thinspace\ldots\ so in order that 
the classical description be valid, the wave packet should remain a wave packet and should 
move according to the laws of classical dynamics. We shall verify that this is so.
\end{quote}
The following calculation is intended to show that such a wave packet always exists. 
Unfortunately, a systematic investigation of different classes of potentials or initial values 
is not performed. Instead, Dirac imposes several conditions for the considered wave 
packets, formulated verbally or in the form of inequalities, which he assumes to be true for 
arbitrary potentials but which need not necessarily be true.  He arrives at the canonical 
equations of motions for the peaks of supposedly arbitrary wave packets.  In 
reality, these conditions impose strong restrictions on the forms of 
initial values and potentials and can be fulfilled only in a few very special cases. 
Quantum-mechanical solutions for the three ``deterministic potentials,'' where 
Ehrenfest's relations agree with NM, are often used to demonstrate so-called ``classical behavior''
of wave packets. It should be borne in mind that this behavior is not generic but 
represents the exception(s)  from the rule.

After the discovery of QT, the community was shocked by the breakdown of NM in 
the microscopic world and it seemed inconceivable that NM should not even survive 
as the classical limit of QT. Schr\"odinger, like Dirac, considered this as evident, and 
wrote at the end of his famous paper about coherent states:\cite{schroedinger:continuous}
\begin{quote}
We can definitely foresee that, in a similar way, wave groups can be constructed which 
move round highly quantized Kepler ellipses and are the representation by wave mechanics
of the hydrogen electron. 
\end{quote}
The coherent states of the harmonic oscillator have been generalized to arbitrary potentials 
in various ways,\cite{zhang_et_al:coherent} but none of these generalizations admits 
a clear transition to the classical (deterministic) limit. Special attention was, of course, devoted 
to the Coulomb potential, but despite intense research, Schr\"odinger's idea could not be realized and 
this chapter has apparently already been closed.\cite{zlatev_et_al:possibility} 

The classical limit of QT is the PHJ, a classical statistical theory  defined by 
Eqs.~\eqref{eq:CA2IH3TMF} and~\eqref{eq:QHKUZ24MF}. The limit 
$\hbar \to 0$ transforms a quantum probabilistic theory into a 
classical probabilistic theory. The behavior of the uncertainty relation 
illustrates this  conclusion in a simple way. For  $\hbar \to 0$ it takes the form  
\begin{equation}
  \label{eq:D3TR12GUPR}
\Delta x \,\Delta p \geq 0
\mbox{,}
\end{equation}
which means that in the classical limit the uncertainty product is in general different 
from zero; a detailed comparison has been reported by Devi and 
Karthik.~\cite{devi_karthik:uncertainity} Almost all states of PHJ will show 
uncertainties; the equality sign in Eq.~\eqref{eq:D3TR12GUPR} just indicates 
that the transition to the deterministic limit (as performed in 
Section~\ref{sec:formal_standard-limit-quant}) is not forbidden. 

The classical limit plays an important role in the prolonged
discussion about the proper interpretation of QT. In the years after the
discovery of QT, a number of dogmas were established, which have 
been repeated since then so many times that they are considered today as 
``well-established'' scientific facts. One of these dogmas states that ``QT provides 
a complete description of individual particles.'' It is hard to understand how a 
probabilistic theory could provide a ``complete'' description of individual events. 
But one should first analyze the possible meanings of the term ``complete.'' A detailed 
analysis shows that this term is ambiguous.\cite{klein:completeness} It may 
mean ``no better theory exists'' (metaphysical completeness) or ``all facts that can 
be observed can be predicted'' (predictive completeness). The still prevailing 
(Copenhagen) standard interpretation claims that QT is complete in both respects.
Einstein, Podolsky, and Rosen (EPR),  showed that QT is predictive-incomplete.~\cite{einstein.podolsky.ea:can} 
In the last paragraph 
of their paper,\cite{einstein.podolsky.ea:can} the authors expressed 
their \emph{belief} that QT is metaphysical-incomplete. EPR's proof of 
predictive incompleteness was correct and could not be attacked, so 
metaphysical incompleteness was attacked instead. The EPR paper was 
misinterpreted as if the  authors had claimed they had proven 
metaphysical incompleteness. Further consequences of this 
misinterpretation are discussed elsewhere.\cite{klein:completeness} The 
important point to note is that metaphysical completeness is a 
philosophical term; physics can test only predictive completeness 
(by comparison with observation). Thus, let us concentrate on the 
question of predictive completeness of QT; in the remaining part 
of this section the term completeness will be used in this sense.

The fact that NM does not emerge from QT, at least in the sense  
of the mathematical limit $\hbar \to 0$, 
presents a painful obstacle to the completeness dogma. Every 
statistical theory, whether classical or quantum, is 
unable to predict individual events and is therefore, by its very 
definition, incomplete. How can quantum theory be complete if 
its classical limit is incomplete? In order to eliminate this problem, 
Bohr created the ``correspondence principle.'' It 
postulates that quantum states become similar to states of NM for 
large values of $S/\hbar$. However, this principle is not in agreement 
with the structure of Schr\"odinger's equation. For large $S/\hbar$ the 
quantum term in Eq.~\eqref{eq:QHKUZ24MF} becomes negligibly 
small and QT becomes similar to PHJ and not to NM; similarity with 
NM requires \emph{in addition} a sharply peaked $\rho$. The 
breakdown of Bohr's correspondence principle in concrete situations 
has been reported many times in the literature; see e.g.\ Cabrera and 
Kiwi~\cite{cabrera_kiwi:large} and Diamond.~\cite{diamond:classically}

Let us finally discuss the logical implication leading from Eq.~\eqref{eq:SPSCHROE}
to Eq.~\eqref{eq:NSEDSCHL}, at the very beginning of the present paper, from the point of 
view of the interpretation of QT.  If Eq.~\eqref{eq:SPSCHROE} really presents a (complete) 
description of a single electron, then the fact that the complete classical 
description, Eq.~\eqref{eq:NSEDSCHL}, follows from Eq.~\eqref{eq:SPSCHROE} 
in the limit $\hbar \to 0$ seems to be evident, not for mathematical but for interpretational 
reasons. But we have mathematically shown that Eq.~\eqref{eq:NSEDSCHL} does 
not follow from Eq.~\eqref{eq:SPSCHROE}. Thus, the premise in the above interpretation of the 
implication ``Eq.~\eqref{eq:SPSCHROE} $\to$ Eq.~\eqref{eq:NSEDSCHL}'' cannot be true, i.e.,
the assumption that Eq.~\eqref{eq:SPSCHROE} provides a (complete) description of a single 
electron cannot be true. Therefore, our result provides  an argument in favor of the 
statistical interpretation,~\cite{ballentine:statistical} which claims that QT provides a 
complete description not of single particles but of statistical ensembles only. 

\section[Conclusion]{Conclusion}
\label{sec:conclusion}

This work has compared the predictions of QT for vanishing $\hbar$
with the predictions of NM. Generally, the predictions of a physical theory are a logical 
consequence of (a) the mathematical form of its basic equations, and (b) the set of initial values. 
Both aspects have been studied in this work, in order to take into account all possible ways to 
obtain predictions of NM from QT in the limit  $\hbar \to 0$. This comparison has been performed 
for the simplest and most significant situation of single particles in external potentials. Our 
conclusion is that NM does not emerge from QT, at least in the sense 
of the straighforward application of the limit $\hbar \to 0$. This
mathematical result should be taken into account in considerations about the interpretation of QT. 

\section{Appendix}
\label{sec:appendix}

Using the theory of generalized functions,\cite{lighthill:generalized} the 
integral equation~\eqref{eq:EFU29EDIREL} can be solved quickly. We discuss 
for simplicity the one-dimensional integral equation,
\begin{equation}
  \label{eq:AODVOT8EE}
f(x)=\int d x\, k(x-y) f(y)
\mbox{,}
\end{equation}
where $k(x)$ is a normalized Gaussian. Introducing the Fourier transforms $F(k)$ and $K(k)$ 
of $f(x)$ and $k(x)$, Eq.~\eqref{eq:AODVOT8EE} takes the form 
\begin{equation}
  \label{eq:FTRVOT78E}
\int d k\, F(k) \left[ 1- \sqrt{2\pi} K(k)\right] =0
\mbox{.}
\end{equation}
Nontrivial solutions for $F(k)$ [and $f(x)$] can exist only if the factor in brackets has zeros. 
This is the case, since the Fourier transform of a normalized Gaussian is again a normalized 
Gaussian and therefore the bracketed quantity vanishes at $k=0$ with a leading quadratic term. Thus, 
the Fourier transform of every solution of~\eqref{eq:AODVOT8EE} must obey  
\begin{equation}
  \label{eq:SHZT45IU9}
F(k)k^{2}=0
\end{equation}
for all $k$. This implies that the Fourier transform of $k^{2}F(k)$, which is proportional to the 
second derivative $f''(x)$ of $f(x)$, vanishes too. This implies $f(x) = a +bx$, in agreement with a 
theorem by Titchmarsh.\cite{titchmarsh:fourier_integrals} [The solution  for $F(k)$ is a linear 
combination of $\delta(k)$ and $\delta'(k)$.]


\begin{thebibliography}{10}

\bibitem{dirac:principles_p88}
P.~A.~M. Dirac,
\textsl{The Principles of Quantum Mechanics}, 4th ed.\
(Oxford University Press, Oxford, 1958), p.\ 88.

\bibitem{allori_zanghi:classical}
V.~Allori and N.~Zanghi,
``On the classical limit of quantum mechanics,''
Found.\ Phys.\ {\bf 39}, 20--32 (2009).

\bibitem{rowe:classical}
E.~G.~P. Rowe,
``Classical limit of quantum mechanics {(}electron in a magnetic
  field{)},''
Am. J. Phys. {\bf 59}, 1111--1117 (1991).

\bibitem{werner:classical}
R.~F. Werner and M.~P.~H. Wolff,
``Classical mechanics as quantum mechanics with infinitesimal $\hbar$,''
Phys. Lett. A {\bf 202}, 155--159 (1995).

\bibitem{rosen:classical_quantum}
N.~Rosen,
``The relation between classical and quantum mechanics,''
Am. J. Phys. {\bf 32}, 597--600 (1964).

\bibitem{schiller:quasiclassical}
R.~Schiller,
``Quasi-classical theory of the nonspinning electron,''
Phys. Rev. {\bf 125}, 1100--1108 (1962).

\bibitem{cohn:quantum}
J.~Cohn,
``Quantum theory in the classical limit,''
Am. J. Phys. {\bf 40}, 463--467 (1972).

\bibitem{gondran:discerned}
M.~Gondran and A.~Gondran,
``The two limits of the {S}chr{\"o}dinger equation in the
  semi-classical approximation,'' in 
\textsl{Foundations of Probability and Physics - 6}, edited by
M. D'Ariano et al. (AIP Conference Proceedings, Melville and New 
York, 2012) p. 318, see also  arXiv:1107.0790 [quant-ph] 

\bibitem{kobe:comments}
D.~H. Kobe,
``Comments on the classical limit of quantum mechanics,''
Am. J. Phys. {\bf 42}, 73--74 (1973).

\bibitem{nikolic:classical}
H.~Nikolic,
``Classical mechanics without determinism,''
Found. Phys. Lett. {\bf 19}, 553--566 (2006).

\bibitem{schroedinger:continuous}
E.~Schr{\"o}dinger,
\textsl{Collected Papers on Wave Mechanics}, 3rd ed.\
(Chelsea Publishing, New York, 1982), p.\ 41.

\bibitem{madelung:quantentheorie}
E.~Madelung,
``Quantentheorie in hydrodynamischer {F}orm,''
Z. Phys. {\bf 40}, 322--326 (1926).

\bibitem{vanvleck:correspondence}
J.~H. {Van Vleck},
``The correspondence principle in the statistical interpretation of
  quantum mechanics,''
Proc. Natl. Acad. Sci. U.S. {\bf 14}, 178--188 (1928).

\bibitem{klein:schroedingers}
U.~Klein,
``{Schr{\"o}dinger's} equation with gauge coupling derived from a
  continuity equation,''
Found. Phys. {\bf 39}, 964 (2009).

\bibitem{kocis:ehrenfest}
L.~Kocis,
``Ehrenfest theorem for the {H}amilton-{J}acobi equation,''
Acta Phys. Polon. A {\bf 102}, 709--716 (2002).

\bibitem{klein:statistical}
U.~Klein,
``The statistical origins of quantum mechanics,''
Physics Research International 808424 (2011), \url{<http://www.hindawi.com/journals/phys/2010/808424.html>}.

\bibitem{ter_haar:problems}
D.~ter Haar,
\textsl{Selected Problems in Quantum Mechanics} (Infosearch Limited, 
London, 1964).

\bibitem{kazandjian:limit}
L.~Kazandjian,
``The $\hbar \to 0$ limit of the {S}chr{\"o}dinger equation,''
Am. J. Phys. {\bf 74}, 557 (2006).

\bibitem{ling:limit}
Song Ling,
``On the $\hbar \to 0$ limit of the {S}chr{\"o}dinger equation,''
J. Chem. Phys. {\bf 96}, 7869--7870 (1992).

\bibitem{home_sengupta:}
D.~Home and S.~Sengupta,
``Classical limit of quantum mechanics,''
Am. J. Phys. {\bf 51}, 265--267 (1983).

\bibitem{titchmarsh:fourier_integrals}
E.~C. Titchmarsh,
\textsl{Introduction to the Theory of Fourier Integrals}, 2nd ed.\ (Oxford, London, 1948),
Theorem 146 on p. 305.

\bibitem{lighthill:generalized}
M.~J. Lighthill,
\textsl{An Introduction to Fourier Analysis and Generalized
  Functions} (Cambridge University Press, Cambridge, 1958).

\bibitem{feynman.hibbs:quantum}
R.~P. Feynman and A.~R. Hibbs,
\textsl{Quantum Mechanics and Path Integrals},
(McGraw-Hill, New York, 1965).

\bibitem{zhang_et_al:coherent}
L.~Zhou and L.~M. Kuang,
``Coherent states: theory and some applications,''
Rev.\ Mod.\ Phys.\ {\bf 62}, 867--927 (1990).

\bibitem{zlatev_et_al:possibility}
I.~Zlatev, W.~Zhang, and D.~H. Feng,
``Possibility that {S}chr{\"o}dinger's conjecture for the hydrogen-atom
  coherent states is not attainable,''
Phys. Rev. A {\bf 50}, R1973--R1975 (1994).

\bibitem{devi_karthik:uncertainity}
A.~R.~Usha Devi and H.~S. Karthik,
``Uncertainty relations in the realm of classical dynamics,''
arXiv:1108.2682 [quant-ph].

\bibitem{klein:completeness}
U.~Klein,
``Is the individuality interpretation of quantum theory wrong ?,''
arXiv:1207.6215 [quant-ph], see also \url{<http://statintquant.net>}.

\bibitem{einstein.podolsky.ea:can}
A.~Einstein, B.~Podolsky, and N.~Rosen,
``Can quantum-mechanical description of physical reality be considered
  complete?,''
Phys. Rev. {\bf 47}, 777--780 (1935).

\bibitem{cabrera_kiwi:large}
G.~G. Cabrera and M.~Kiwi,
``Large quantum-number states and the correspondence principle,''
Phys. Rev. A {\bf 36}, 2995--2998 (1987).

\bibitem{diamond:classically}
J.~J. Diamond,
``Classically forbidden behavior of the quantum harmonic oscillator for
  large quantum numbers,''
Am. J. Phys. {\bf 60}, 912--916 (1992).

\bibitem{ballentine:statistical}
L.~E. Ballentine,
``The statistical interpretation of quantum mechanics,''
Rev.\ Mod.\ Phys.\ {\bf 42}, 358--381 (1970).

\end{thebibliography}
\end{document}